\documentclass[%
groupedaddress,
preprint,
 amsmath,amssymb,
 aps,
pra,
]{revtex4-1}

\usepackage{graphicx}
\usepackage{dcolumn}
\usepackage{bm}

\begin{document}

\newcommand{\be}{\begin{equation}}
\newcommand{\ee}[1]{\label{#1}\end{equation}}
\newcommand{\bem}{\begin{eqnarray}}
\newcommand{\eem}[1]{\label{#1}\end{eqnarray}}
\newcommand{\eq}[1]{Eq.~(\ref{#1})}
\newcommand{\Eq}[1]{Equation~(\ref{#1})}
\newcommand{\vp}[2]{[\mathbf{#1} \times \mathbf{#2}]}


\title{ Motion of vortices in ferromagnetic spin-1 BEC}

\author{E.  B. Sonin}
 \affiliation{Racah Institute of Physics, Hebrew University of
Jerusalem, Givat Ram, Jerusalem 91904, Israel}

\date{\today}

\begin{abstract}
The paper investigates dynamics of nonsingular vortices  in a ferromagnetic spin-1 BEC, where  spin and mass superfluidity coexist in the presence of uniaxial anisotropy (linear and quadratic Zeeman effect). The analysis is based on  hydrodynamics following from  the Gross--Pitaevskii theory. Cores of  nonsingular vortices are skyrmions with charge, which is tuned  by uniaxial anisotropy and can have any fractal value between 0 and 1. There are circulations of mass and spin currents around these vortices.
The results are compared with the equation of vortex motion derived earlier in the Landau--Lifshitz--Gilbert theory for magnetic vortices in easy-plane ferromagnetic insulators. In the both cases the transverse gyrotropic force (analog of the Magnus force in superfluid and classical hydrodynamics) is proportional to the charge of skyrmions in vortex cores.
\end{abstract}

\maketitle


\section{Introduction}

Similarly to superfluid $^3$He, the ferromagnetic spin-1 BEC  combines properties of a common  superfluid and of a magnetically ordered system  \cite{UedaR}. Correspondingly, one may expect coexistence and interplay of spin superfluidity \cite{Adv,Mac} and more common mass superfluidity.  Like common ferromagnets, the spin-1 BEC is   described at macroscopical scales by the Landau--Lifshitz--Gilbert (LLG) theory \cite{LLstPh2} but extended by inclusion of an additional degree of freedom of fluid motion as a whole \cite{Lam08,Sp1}.

Emergence of superfluidity is conditioned by special topology of the order parameter space.   In the case of scalar superfluids the order parameter space is a circumference on a complex plane of a complex wave function. Topology of the circumference allows mass superfluidity, since current states map on paths winding around the circumference. These mappings cannot be reduced to a point by continuous transformation without leaving the circumference. As for spin superfluidity in magnetically ordered media, this requires the easy-plane anisotropy in the spin space \cite{Adv,Mac}. This anisotropy can emerge not only from crystal anisotropy, but also from long-range magnetostatic (dipolar) interaction as shown for the magnon condensate in  yttrium-iron-garnet  magnetic films  \cite{Son17}.

Manifestation of superfluidity is macroscopic persistent currents proportional to gradients of phase (phase of the wave function for mass superfluidity and the angle of spin rotation around some axis in the case of spin superfluidity). Persistent current states are metastable states, but they  lose stability when phase gradients reach some critical values. After this frequent phase slips destroy persistent currents and relaxation  to current-less ground states occurs. Vortices also emerge in the  equilibrium  rotating superfluids.
  Thus vortices are crucial for the phenomenon of superfluidity. Its very existence as linear topological defects requires the same topology as necessary for existence of superfluidity.   
  
Investigation of vortices in scalar superfluids started  from the seminal works  of Onsager \cite{Ons} and Feynman \cite{Fey}.  In magnetically ordered systems magnetic vortices  also were known long ago. Magnetic vortex is an example of topological defects in magnetically ordered solids, which were in the focus of scientific activity of Arnold Markovich Kosevich and his colleagues \cite{KVM,KIK,KIKr}. At motion of a magnetic vortex a reactive gyrotropic force proportional and normal to its velocity emerges. This gyrotropic force was first revealed by Thiele \cite{thiele} for magnetic bubbles in ferromagnetic films. Later the gyroscopic force was derived for magnetic vortices in easy-plane magnets \cite{huber,NikSon}.  In contrast to friction force also proportional to the vortex velocity, the gyrotropic force does not depend on  the spin texture inside the vortex core, but does depend on circulation of the spin phase (the angle of spin rotation in the easy-plane). 

The goal of the present paper is derivation of the equation of motion of a nonsingular vortex in the ferromagnetic spin-1 BEC. Nonsingular vortices are possible in multi-component superfluids when vorticity is not concentrated at a singular line  (axis of the vortex) but is continuously distributed over a core of finite radius.  First they were revealed in the $A$ phase of superfluid $^3$He \cite{Sal85}. The energy of nonsingular vortices is smaller than of singular ones, and phase slips with nonsingular vortices are more probable.   In the ferromagnetic spin-1 BEC mass and spin superfluidity coexist, and a nonsingular  vortex is a hydrodynamic and a magnetic vortex at the same time,  i.e., it has circulations of mass and spin currents around it. This has an impact on phase slips destroying mass and spin supercurrent \cite{Sp1}. Normally the core radius of nonsingular vortices  exceeds microscopical scales, and the hydrodynamical approach is sufficient for derivation of the equation of vortex motion.

There are two methods to derive the equation of vortex motion. The first one is using the solvability condition. The hydrodynamic equations (the LLG equations in the case of magnetic vortices)  are linearized with respect to small perturbations of the static solution for a resting vortex. Perturbations are produced by vortex motion and by  currents past the vortex.  This yields nonuniform linear equations. It is not necessary to solve the equations explicitly. The equation of vortex motion is derived from a condition for their solvability, which is called also the condition of the absence of secular terms. This approach was used in the past for the analysis of dynamics of nonsingular vortices in $^3$He-$A$ \cite{Son86}. The second method uses  the conservation law for momentum in a Galilean invariant medium, or for quasimomentum if Galilean invariance is broken following from Noether's theorem for translationally invariant media.  In the absence of external forces  or friction force, which violate the conservation law for (quasi)momentum, the second method  yields exactly the same equation of vortex motion as the first method \cite{NikSon}. Here we use the second method. The equation of motion is the balance equation of  forces on the vortex. There are a gyrotropic force proportional to the vortex velocity (analog of the Magnus force on a hydrodynamic vortex) and a force proportional to mass and spin supercurrents past the vortex (Lorentz force in superfluid hydrodynamics).  These forces depend on topological charges of vortices but not  on details of the core structure, in contrast to the friction force, which does depend on details  of the vortex core but is not investigated in the present work. 
  
 We start our analysis from formulation of hydrodynamics of  ferromagnetic spin-1 BEC following from the  Gross--Pitaevskii theory (Sec.~\ref{GP}).   Section \ref{vdL} reviews dynamics of magnetic vortices in ferromagnet insulators, where mass (charge) currents are absent. This is necessary  for comparison with dynamics of vortices in the ferromagnetic spin-1 BEC, where both mass and spin currents are possible.  Sections~\ref{GP} and  \ref{vdL} review previously known results.  Dynamics of vortices in the ferromagnetic spin-1 BEC is addressed in
 Sec.~\ref{vdb}. Discussion and conclusions are presented in Sec.~\ref{dis}.
 
   \section{Gross--Pitaevskii theory for  ferromagnetic spin-1 BEC} \label{GP}

In  the spin-1 ferromagnetic BEC  the condensate wave function (the order parameter) can be presented as  a 3D  complex vector in the spin space:
\be
\bm \psi ={\psi_0 \over \sqrt{2}} (\bm m+i \bm n),
        \ee{PW}  
where scalar  $\psi_0$ and two unit mutually orthogonal vectors $\bm m$ and $\bm n$ are real. The two unit vectors $\bm m$ and $\bm n$ together with the third vector 
\be
\bm s=\bm m \times \bm n
        \ee{}
 form a triad of three real orthogonal unit vectors.  The unit vector $\bm s$ points out direction of full spin polarization. It is an analog of the orbital vector $\bm l$ in the $A$ phase of superfluid $3$He, which shows direction of the orbital moment of Cooper pairs.
Neutral and charged superfluids with such order parameter are called chiral or $p_x+ip_y$ superfluids.

The gauge transformation of the ferromagnetic spin-1 order parameter,
\bem \bm m+i \bm n ~\to~(\bm m+i \bm n)e^{i\theta}
= (\bm m \cos\theta- \bm n\sin\theta )+i(\bm m \sin\theta+ \bm n\cos\theta ),
       \eem{}
 is equivalent to rotation around the axis $\bm s$  by the angle $ \phi_s=-\theta$ and therefore is not an independent symmetry transformation. So the full point symmetry group of the order parameter is the group $SO(3)$ of three-dimensional rotations. The group is not abelian, and the angle of rotation around any axis including the axis $\bm s$ depends on the path along which the transformation is performed. In particular, if we deal with the phase $\theta=-\phi_s$, a result of two small consecutive variations  $\delta_1$ and $\delta_2$ of $\theta$ depends on the order of their realizations:
 \be
 \delta_1 \delta_2 \theta-\delta_1 \delta_2 \theta=\bm s\cdot [\delta_1 \bm s \times \delta_2 \bm s].
              \ee{com12}
This means that the phase $\theta$ is not well defined globally, although its infinitesimal variations still make sense, and the quantum-mechanical definition of the superfluid velocity, 
\be
\bm v_s={\hbar\over m}\bm \nabla \theta,
                                               \ee{9.1}
is valid. Here $m $ is the mass of a boson. Because of \eq{com12} variation of the superfluid velocity is determined not only by  variation of the phase $\theta$ itself but also by variation of the spin vector $\bm s$. As a result,  the superfluid velocity is not curl-free. Replacing  $\delta_1$ and $\delta_2$ in \eq{com12} with two gradients $\nabla_1$ and $\nabla_2$ along two different directions ($x$ and $y$,  or $y$ and $z$, or $z$ and $x$),  \eq{com12} yields the Mermin--Ho relation  \cite{Mer} between vorticity and spatial variation of $\bm s$: 
\be
\bm \nabla \times \bm v_s={\hbar\over 2m}\epsilon_{ikn} s_i \bm \nabla s_k \times \bm \nabla s_n.
                                              \ee{9.2}
This relation has a dramatic impact on hydrodynamics of chiral superfluids.

For bosons with spin 1 the most general  Lagrangian of the Gross--Pitaevskii  theory is 
\be
{\cal L}={i\hbar \over 2}\left(\bm \psi^*\cdot {\partial \bm \psi \over \partial t}-\bm \psi \cdot {\partial \bm \psi^* \over \partial t} \right) - H,
    \ee{Lag3}
where $H $ is the Hamiltonian, which can depend on the wave function $\bm \psi$ and its gradients.

According to Noether's theorem, gauge invariance leads to the mass continuity equation:
 \be
 {\partial \rho\over \partial t}+\bm\nabla\cdot \bm j=0,
     \ee{mC}
where 
\be
\rho={1\over i\hbar }\left(  {\partial {\cal L} \over \partial \dot {\bm \psi}}\cdot\bm \psi -  {\partial {\cal L} \over \partial \dot {\bm\psi ^*}}\cdot\bm\psi^* \right) =m\bm \psi^*\cdot \bm \psi
    \ee{}
is the mass density and 
\bem 
\bm j ={1\over i\hbar }\left( {\partial {\cal L} \over  \partial \bm \nabla \psi_j} \psi_j -  {\partial {\cal L} \over \partial \bm \nabla \psi _j^*} \psi_j^*\right)
   \eem{}      
is  the mass current.
  
Noether's theorem  connects translational invariance with the conservation law 
\be
{\partial g_i\over \partial t}+ \nabla_j\Pi_{ij}=0,
     \ee{cur-g}
where
\be
\bm g  =-{\partial {\cal L} \over \partial \dot \psi_j}\bm \nabla\psi_j -  {\partial {\cal L} \over \partial \dot \psi _j^*}\bm \nabla\psi_j^*  
=-{i\hbar \over 2}(\psi_j^*\bm  \nabla  \psi_j-\psi_j\bm \nabla \psi_j^*)
      \ee{quaC}
is a current, which can be different from the mass current in general, and 
\bem
\Pi_{ij}=- \nabla_i \psi_k {\partial {\cal L} \over \partial \nabla_j  \psi_k}
- \nabla_i \psi_k^* {\partial {\cal L} \over \partial \nabla_j  \psi_k^*} +\delta_{ij} {\cal L}
  \eem{quaM}
is some flux tensor. 

The third conservation law follows from Noether's theorem if the Hamiltonian is invariant with respect to any rotation in the spin space:
 \be    
         {\partial  S_i  \over \partial t}+\nabla_jJ_{ij}   =0,  
    \ee{SpCon}
where 
\be 
\bm S   =- {\partial {\cal L} \over \partial \dot {\bm \psi}}\times \bm \psi -  {\partial {\cal L} \over \partial \dot  {\bm\psi ^*}} \times   \bm\psi^*  =i\hbar [\bm \psi \times \bm \psi^*]={\hbar\rho\over m}\bm s
   \ee{}
is the spin density and 
\be
J_{ij}   =-\epsilon_{ijk}\left( {\partial {\cal L} \over \partial \nabla_j \bm \psi}\cdot \nabla_k \bm \psi+  {\partial {\cal L} \over \partial \nabla_j  \bm\psi ^*} \cdot\nabla_k\bm \psi^*\right)
      \ee{SC}
is the spin current tensor.

If the BEC  is Galilean invariant as it should be in the absence of optical lattices, the Hamiltonian  and the nonlinear Schr\"odinger equation are
\bem
H ={\hbar^2 \over 2m}\nabla_i\psi_j^* \nabla_i\psi_j+{V|\bm \psi|^4\over 2},
      \eem{Ham3}
\be
i\hbar {\partial \bm \psi \over \partial t}={\delta  H\over \delta \bm \psi^*}=-{\hbar^2  \nabla_j^2\bm \psi \over 2m}+V|\bm \psi|^2\psi.
    \ee{EqHam} 
Only for a Galilean invariant superfluid the current $\bm g$ coincides with the mass current $\bm j$, which at the same time is the momentum density of the superfluid. Then the conservation law (\ref{cur-g}) is the conservation of the momentum, and the flux tensor
\bem
\Pi_{ij}= {\hbar^2\over 2M}\left(\nabla_i \psi_k \nabla_j \psi_k^* +\nabla_i \psi_k^* \nabla_j \psi\right) +\delta_{ij} P,
  \eem{}
is the momentum flux tensor with the  pressure  given by
\be
P={\cal L}={V|\bm \psi|^4\over 2}-{\hbar^2\over 4m}  \nabla^2|\bm \psi|^2 .
   \ee{}
In the absence of Galilean invariance we shall call the current $\bm g$ the quasimomentum density and the tensor $\Pi_{ij}$ the quasimomentum flux tensor. If the superfluid is in  a periodic potential  (BEC in an optical lattice, e.g.) the current $\bm g$ is a density of the quasimomentum indeed as it is defined in the Bloch band theory \cite{EBS}.  The Gross--Pitaevskii theory for $p_x+ip_y$ superfluids, which is presented here, has already been used in the past for the $A$ phase of superfluid $^3$He \cite{Son84,EBS}.

Transition from the Gross--Pitaevskii theory to the hydrodynamical description is realized by the generalized Madelung transformation.  
After the transformation the superfluid  is described by the mass density $\rho=m \psi_0^2$, the orbital vector $\bm s$, and the quantum-mechanical phase $\theta$.  In the hydrodynamical approach usually they neglect dependence of the energy on density gradients (gradients of $\psi_0$) responsible for quantum pressure \cite{EBS}. The Lagrangian and the Hamiltonian  after the Madelung transformation  become 
\be
{\cal L}= -{\hbar \over m} \rho {\partial \theta \over \partial t}
 - H,
    \ee{LagM}      
 \bem
H = {\rho \over 2} v_s ^2+   {\hbar^2 \rho \over 4 m^2 }\nabla_i \bm s\cdot \nabla_i \bm s  +{V\rho^2\over 2m^2}.
      \eem{hamIn}
In hydrodynamics two canonical equations of motion are the continuity equation  (\ref{mC}) and the Josephson equation for the phase $\theta$,
\be 
{\hbar \over m}{\partial \theta \over \partial t}+ \mu_0+{v_s^2\over 2}=0.
     \ee{JE}
They are similar to those in a non-chiral superfluid. Here 
\be
\mu_0 = {\hbar^2  \over 4 m^2 }\nabla_i \bm s\cdot \nabla_i \bm s  +V\rho
      \ee{mu}
is the chemical potential of the superfluid at rest. 

The third hydrodynamical equation  is for the unit vector $\bm s$:
 \be    
         {\partial \bm s  \over \partial t}+ (\bm v_s\cdot   \bm \nabla) \bm s -{\hbar  \over 2 m \rho }\left[ \bm s \times      \nabla _i(\rho \nabla_i \bm s)\right]     =0.  
    \ee{orbGPa}
For a fluid at rest ($\bm v_s=0$) \eq{orbGPa} is identical to the LLG  equation for magnetization in a solid ferromagnetic insulator. 
 
After the Madelung transformation the  spin current tensor (\ref{SC})  becomes 
\bem
 J_{ij} =  S_i   v_{sj }   -\epsilon_{ikl}s_k {\partial H\over \partial \nabla_j s_l} 
 = S_i   v_{sj }   -{\hbar^2\rho  \over 2 m^2 }\epsilon_{ikl}s_k \nabla_j s_l.
       \eem{spCur}     
The first term in the expression   for the spin current presents advection of spin by fluid motion as a whole. This effect is trivial and has nothing to do with special conditions required for existence of spin superfluidity. Only the second term,
\be 
 j_{ij} =     -{\hbar^2\rho  \over 2 m^2 }\epsilon_{ikl}s_k \nabla_j s_l,
       \ee{spCurS}     
connected with stiffness of the spin texture will be later called the spin supercurrent. 

The Euler equation for the velocity $\bm v_s$ must follow from the Josephson equation (\ref{JE})  by applying  the gradient operator.
But one should take into account  non-commutativity of the operators $\partial /\partial t$ and $\bm \nabla$ at their actions on the phase $\theta$. Namely, according to \eq{com12} 
 \be
\nabla_i  {\partial \theta\over\partial t}- {\partial (\nabla_i\theta)\over\partial t}
=\nabla_i  {\partial \theta\over\partial t}-{m\over \hbar} {\partial v_{si}\over\partial t}=\bm s\cdot \left[\nabla_i  \bm s \times {\partial \bm s\over\partial t} \right].
              \ee{}
After some algebra using the Mermin--Ho relation (\ref{9.2}) and the  equation (\ref{orbGPa}) of spin dynamics one obtains the Euler equation 
\be
\dot {\bm v}_s+( \bm v_s \cdot \bm  \nabla) \bm v_s
+  \bm \nabla\mu_0+  {\hbar ^2  \over 2 m^2}\bm \nabla s_i {\nabla_j (\rho \nabla_j s_i )\over \rho}=0.
  \ee{}

\begin{figure}[t]
\includegraphics[width=.5\textwidth]{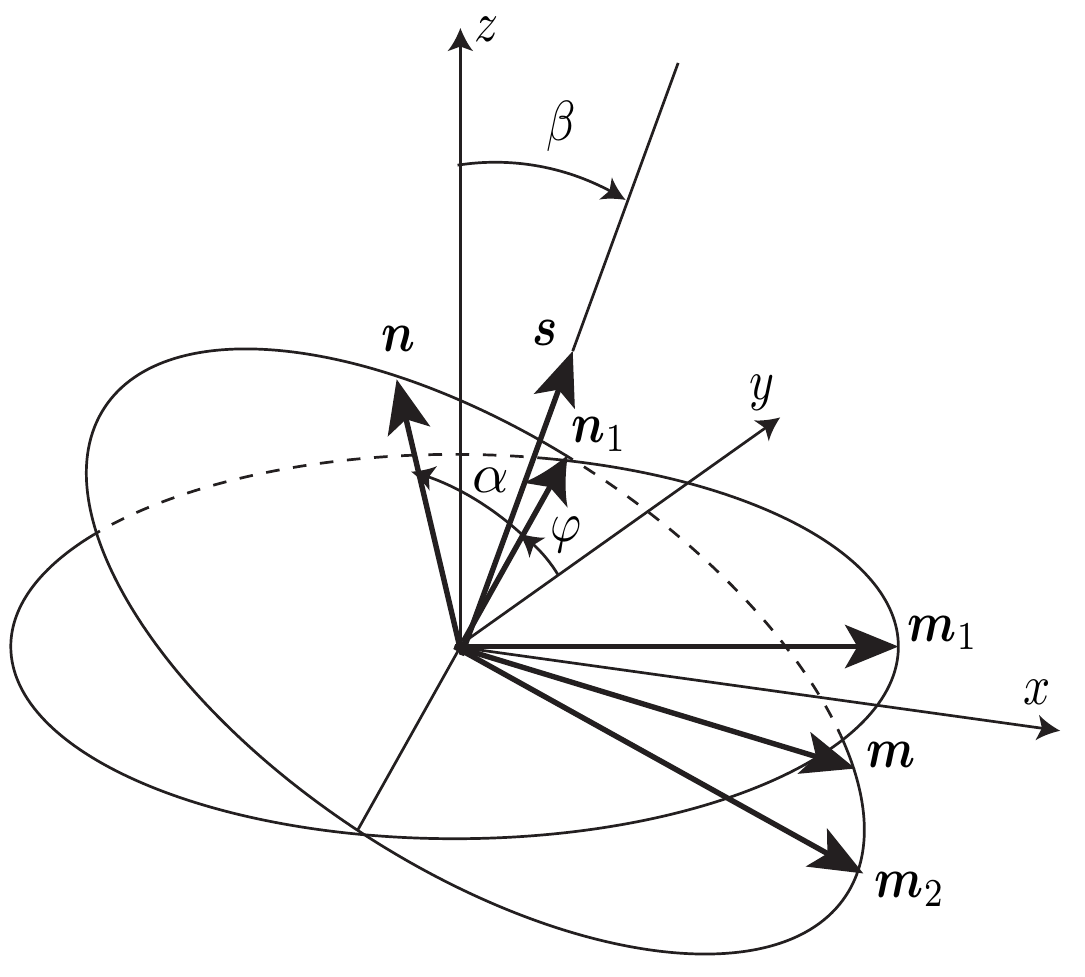}
\caption[]{Euler angles for the wave function triad. The original positions of $\bm m$, $\bm n$, and $\bm s$  are along the axes $x$, $y$, and $z$ respectively. The first rotation by the angle $\varphi$ is in the plane $xy$, which brings the first two vectors to the positions $\bm m_1$ and $\bm n_1$. The second rotation by the angle $\beta$ is in the plane confining the axis $z$ and the vector $\bm m_1$ (around the vector $\bm n_1$). This brings the vector $\bm s$ to its final position and rotates the vector $\bm m_1$ to $\bm m_2$. The last third rotation by the angle $\alpha$ is around the vector $\bm s$, which transforms the vectors $\bm m_2$ and $\bm n_1$ to the final vectors $\bm m$ and $\bm n$ determined by \eq{tria}.}
\label{Fig1}
\end{figure}

It is possible to avoid  dealing with the globally undefined phase $\theta$  by introducing Euler angles as hydrodynamical variables. They determine  rotation of the triad  $\bm m,\bm n,\bm s$ with respect to the original triad $\hat x,\hat y, \hat z$ as shown in Fig.~\ref{Fig1}:
\be
\begin{array} {ccc}
m_x =\cos \beta \cos \alpha \cos \varphi-\sin \alpha \sin\varphi, & m_y =\cos \beta \cos \alpha \sin \varphi+\sin \alpha \cos \varphi, & m_z=-\sin\beta \cos \alpha   \\
n_x=-\cos \beta \sin\alpha \cos \varphi -\cos \alpha \sin \varphi,& n_y =-\cos \beta \sin\alpha \sin \varphi+\cos \alpha\cos\varphi ,& n_z=\sin \beta \sin\alpha \\
s_x=\sin\beta \cos \varphi, & s_y =\sin\beta \sin \varphi, & s_z =\cos \beta
\end{array}
    \ee{tria}
In terms of the Euler angles the superfluid velocity is
\be
\bm v_s=-{\hbar\over m}(\bm \nabla \alpha+\cos \beta \bm \nabla \varphi),
   \ee{vEu}
while the Mermin--Ho relation becomes
\be
\left[\bm \nabla \times \bm v_s\right]={\hbar \over m}\sin\beta[ \bm \nabla \beta  \times \bm \nabla \varphi  ].  
  \ee{}
According to \eq{spCurS}, the current of the $z$-component of spin is
\be
\bm j_z =-{\hbar^2 \rho\over 2m^2}\sin^2\beta  \bm \nabla\varphi.
    \ee{}
Using the Euler  angles as variables the momentum flux tensor is
\be
\Pi_{ij}=\rho v_iv_j+  {\hbar ^2 \rho \over 2 m^2}\left(\nabla_i \beta \nabla_j \beta+\sin^2 \beta  \nabla_i \varphi \nabla_j \varphi\right)+P\delta_{ij},
    \ee{}
where the pressure $P$ is equal to the Lagrangian determined by \eq{LagM}.

Up to now our equations were isotropic in the spin space of the vector $\bm s$. But in an isotropic ferromagnet  vortices as stable linear topological defects do not exist, as well as neither mass nor spin superfluidity is possible .   Thus we shall add to our Hamiltonian terms breaking spherical symmetry but still invariant with respect to rotations around the axis $z$ (uniaxial anisotropy):
\be
H_A = - \gamma H_{ef} S s_z + {\rho Gs_z^2\over 2}.
         \ee{H_A}
Here $\gamma$ is the gyromagnetic ratio. The first term linear in $s_z$ is the Zeeman energy. The field $H_{ef}$ can be an external magnetic field but not necessarily. Processes violating the conservation law of spin usually are weak in comparison with the exchange interaction. Pumping magnons one can create a non-equilibrium  $z$ component of spin, which relaxes quite slowly, and this relaxation can be compensated by continuing magnon  pumping. With good accuracy one may consider  this state as a quasi-equilibrium state with fixed $z$ component of spin. Such states under the name magnon BEC were realized both in solids \cite{Dem6} and in ferromagnetic spin-1 BEC \cite{MagBEC}. Then $H_{ef}$ is a Lagrange multiplier, which determines the value of fixed total spin. The second term in \eq{H_A} is called in magnetism  easy-axis ($G<0$) or easy-plane ($G>0$) anisotropy. In the theory of cold atoms they call it the quadratic Zeeman energy \cite{UedaR}. The anisotropy energy determines two possible phases with the orientational phase transition between them. At  $\gamma SH_{ef} > \rho G$ the energy is minimal at $s_z=1$ (easy-axis phase), while at $\gamma SH_{ef} < \rho  G$ the spin is confined in the plane parallel to the $xy$ plane and corresponding to $s_z=\gamma SH_{ef} /\rho G$ (easy-plane phase). Since in the easy-plane phase invariance with respect to rotations around the axis $z$ is spontaneously broken, it is called also broken-axisymmetry phase \cite{UedaR}.

Further we consider the case of incompressible liquid, when it is  enough to analyze only soft spin modes and to neglect density variation. 
Using the Euler angles  for the unit vector $\bm s$ as in \eq{tria},
the spin Hamiltonian including the anisotropy terms is 
\bem
H=  \rho  \left\{{v_s^2\over 2} + {\hbar ^2 \over  4m^2}\left[\sin^2\beta (\nabla \varphi)^2 +(\nabla \beta)^2\right]
+{G(\cos\beta -s_0)^2\over 2}\right\}
 \nonumber \\ 
=  {\rho\hbar ^2 \over  m^2}   \left[{(\bm \nabla \alpha+\cos \beta \bm \nabla \varphi)^2\over 2} +{\sin^2\beta (\nabla \varphi)^2 +(\nabla \beta)^2\over 4}+{m^2G(\cos\beta -s_0)^2\over 2\hbar^2}\right],
   \eem{Hpol}
where 
\be
s_0={\gamma SH_{ef} \over \rho G}={\gamma \hbar H_{ef} \over m G},
    \ee{s0}
and the superfluid velocity satisfies the incompressibility condition
\be
\bm \nabla\cdot \bm v_s =- \bm \nabla\cdot (\bm \nabla \alpha+\cos \beta \bm \nabla \varphi)=0.
     \ee{}   
  The equations of spin dynamics in polar angles are
\be 
\dot \beta  +(\bm v_s\cdot \bm \nabla)  \beta={\hbar \sin \beta \over  2m } \nabla ^2 \varphi  -{\hbar  \cos \beta \over  m }\bm   \nabla  \varphi \cdot \bm \nabla \beta , 
     \ee{sde1}
\bem
\dot \varphi+(\bm v_s\cdot \bm \nabla)\varphi=  -{\hbar \over  2m}\left[ (\bm \nabla \varphi)^2\cos\beta -{\nabla ^2 \beta \over \sin\beta}
\right]
+{mG(\cos\beta -s_0) \over \hbar}.
         \eem{sde2}

\section{Dynamics of vortices in the LLG theory for localized spins} \label{vdL}

Although our final goal is dynamics of vortices in the ferromagnetic spin-1 BEC, it is useful for later discussion and comparison to start from magnetic vortices in ferromagnetic insulators, where spin carriers are localized and the degree of freedom of motion of the medium  as a whole is absent. The original LLG theory referred  exactly to this case.  The Lagrangian and the Hamiltonian  of the LLG theory in angle variables are 
\be
{\cal L}= {\hbar \over m} \rho_1 \cos \beta {\partial \varphi \over \partial t}
 - H,
    \ee{LagMl}      
 \bem
H=    {\hbar ^2 \rho_2 \over  4m^2}\left[\sin^2\beta (\nabla \varphi)^2 +(\nabla \beta)^2\right]
+{\rho_2 G(\cos\beta -s_0)^2\over 2}.
   \eem{HpolL}
The equations of spin dynamics are
\be 
\dot \beta ={\hbar \rho_2 \sin \beta \over  2m\rho_1 } \nabla ^2 \varphi  -{\hbar \rho_2 \cos \beta \over  m\rho_1 }\bm   \nabla  \varphi \cdot \bm \nabla \beta , 
     \ee{sd1}
\bem
\dot \varphi=  -{\hbar\rho_2 \over  2m\rho_1}\left[ (\bm \nabla \varphi)^2\cos\beta -{\rho_2\nabla ^2 \beta \over \sin\beta}
\right]
+{m\rho_2G(\cos\beta -s_0) \over \hbar\rho_1}.
         \eem{sd2}
Here we introduced the densities $\rho_1$ and $\rho_2$. The density $\rho_1$ in the first term of the Lagrangian (the Wess--Zumino term)  determines the spin density $S = \hbar\rho_1 /m$, which is a constant in the LLG theory. The second density $\rho_2$ determines phase stiffness of the magnetic order parameter.  In the Heisenberg model of ferromagnetic insulators $\rho_2$ is proportional to  the exchange interaction between spins at neighboring sites. Introduction of the densities $\rho_1$ and $\rho_2$ makes comparison of the LLG theory for ferromagnetic insulators and spin-1 BEC more convenient: if $\rho_1$ and  $\rho_2$ are equal to the total mass density $\rho$ these  equations coincide with Eqs.~(\ref{sde1}) and (\ref{sde2}) but without $\bm v_s$-dependent terms.

\begin{figure}[t]
\includegraphics[width=0.7\textwidth]{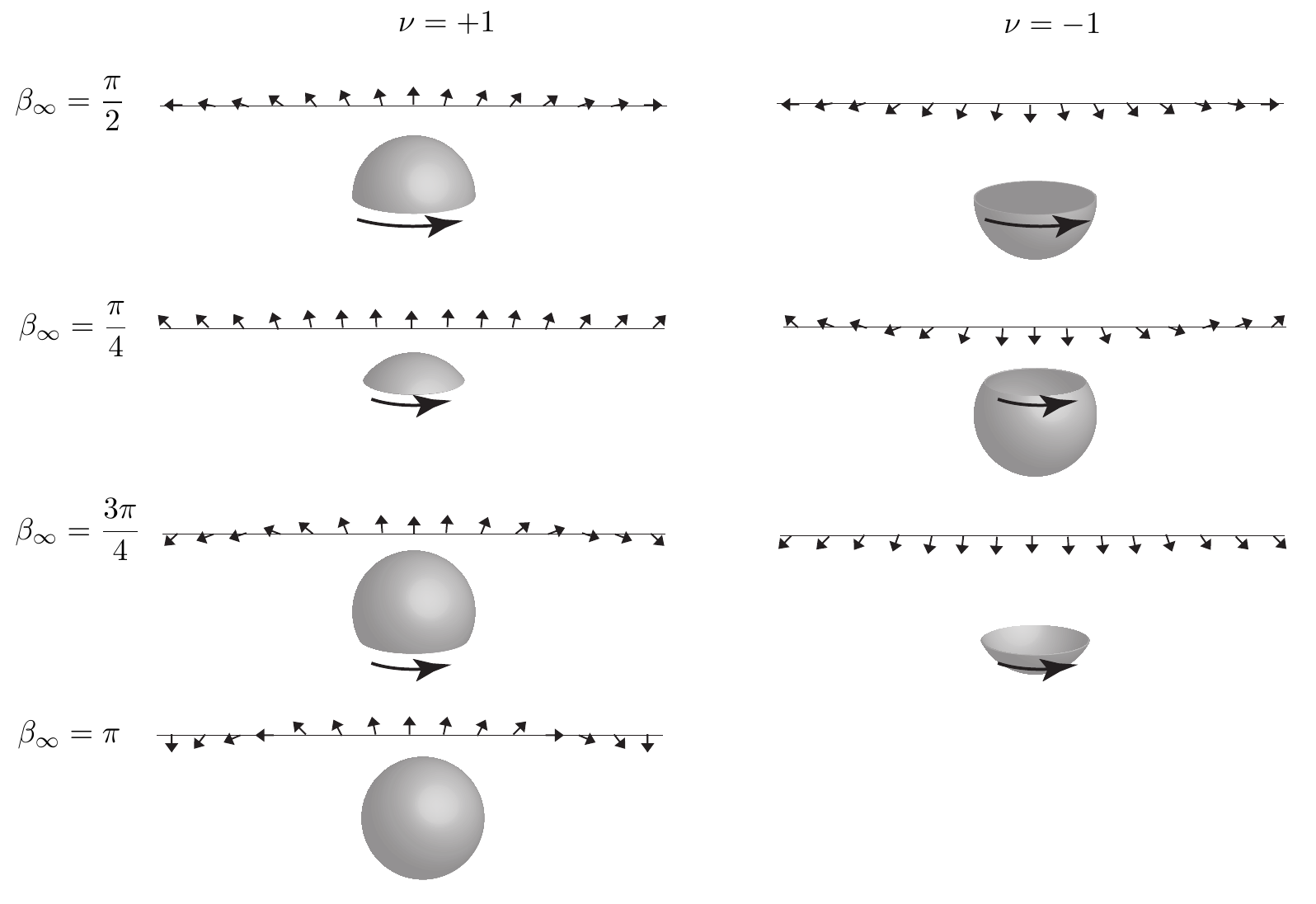}
\caption[]{ Spin vectors  $\bm s$ in axial cross-sections of skyrmion cores and mapping  on the space $S_2$ for vortex states with polarizations $\nu =\pm1$ and polar angles $\beta_\infty =\pi/4,~\pi/2,~3\pi/4$, and $\pi$. Larger arrows show direction of circular spin currents around the vortex (skyrmion) axis.}
\label{Fig3}
\end{figure}

Without anisotropy the order parameter space is $S_2$, which is  a 2D surface of a unit  sphere in the 3D space. Every point of $S_2$ corresponds to some direction of the vector $\bm s$.  Spin superfluidity and vortices are possible  only  in the easy-plane phase when the order parameter space reduces to a circumference of the sphere $S_2$ corresponding to some fixed value of $s_z$ ($|s_z|<1$). But only periphery of the vortex very far from its axis maps on this circumference. The core of the vortex maps on an upper (northern) or lower (southern) part of the sphere. The vortex is characterized by two topological numbers \cite{NikSon}. The first one is the winding number, i.e., the number of rotations the spin makes on going around a vortex (the analogue of the number of circulation quanta for a vortex in superfluid hydrodynamics). The second number, which can be called polarization, takes two values $\nu=\pm 1$. Two signs  correspond to a sign of the spin component $s_z$ at the vortex axis. We choose direction of the axis $z$ so that the in-plane spin component rotates counterclockwise around it.  The positive polarization corresponds to mapping on the northern part of the sphere, while the negative polarization points out mapping on the southern one. Mapping of vortex states with two polarizations and various values of the polar angle $\beta_\infty$ far from the vortex  are shown in Fig.~\ref{Fig3}. The vortex core at $\beta_\infty=\pi$ has the structure of a skyrmion with the charge $Q=1$. The skyrmion charge is a measure of wrapping of the spin vector around the sphere $S_2$ and equal to $Q=\sin^2 {\beta_\infty\over 2}$. At $\beta_\infty=\pi/2$ when at periphery the spin is confined in the $xy$ plane, the core skyrmion  is a meron, or a half-skyrmion with  the skyrmion charge one-half \cite{braun}. Other values of $\beta_\infty$ correspond to other fractional skyrmion charges. Thus in the presence of uniaxial anisotropy the skyrmion charge is not quantized and may vary continuously. 

Skyrmions shown in Fig.~\ref{Fig3} are Neel skyrmions with non-zero magnetostatic charges proportional to $\bm \nabla\cdot \bm s$. But rotation in the spin space around the axis $z$ transforms skyrmions to Bloch skyrmions. Our model is invariant with respect to this rotation since it ignores the magnetostatic  interaction.

In a straight axisymmetric vortex spin depends only on two polar coordinates $r,\phi$. For a single-quantum vortex $\varphi$  does not depend on  $r$ and is equal to the azimuthal angle $\phi$. The gradient of the spin phase $\varphi$, 
\be
\bm \nabla \varphi={[\hat z \times \bm r] \over r^2},
       \ee{grVor}
has only the azimuthal component  $\propto 1/ r$.
The polar angle $\beta$ depends only on $r$. Then the Hamiltonian  (\ref{HpolL}) does not depend on the angle $ \phi$ and becomes
\be
 H=  \rho_2  \left\{{\hbar ^2 \over  4m^2}\left[{\sin^2\beta \over r^2} +\left(d \beta\over dr\right)^2\right]
+{G(\cos\beta -s_0)^2\over 2}\right\}.
   \ee{HpolA}
The Euler--Lagrange equation for this Hamiltonian describes spin texture in a resting vortex:
 \bem
{d ^2 \beta \over dr^2}+{1\over r}{d  \beta \over dr}-\sin\beta\left({ \cos\beta\over r^2} 
-{\cos\beta -s_0 \over \xi^2}\right)=0,
         \eem{beta}
 where
 \be
 \xi={\hbar \over m\sqrt{2G}},
      \ee{xi} 
and $s_0 = \cos \beta_\infty$ is the value of $s_z$ at large distances from the vortex axis. At small $r$  $\beta \propto r$, while at large $r$ $\beta$ approaches to the equilibrium value $\beta_\infty$:
\be
\beta \approx \beta_\infty - {\xi^2\cos \beta_\infty \over r^2 \sin \beta_\infty}.
    \ee{} 
One can define the core radius as a distance $r$ at which the correction to the asymptotic  value $\beta_\infty$ becomes comparable with $\beta_\infty$ itself. This yields the core radius of the order $r_c \sim \xi$  excepting  very small 
$\beta_\infty $, when  \eq{beta} becomes
\bem
{d ^2 \beta \over dr^2}+{1\over r}{d  \beta \over dr}-{ \beta\over r^2} 
-{(\beta^2 -\beta_\infty^2)\beta \over 2\xi^2}=0.
         \eem{beta1}
This equation is identical to the Gross-Pitaevskii equation for radial distribution of the density of the vortex in a single-component superfluid. It determines  the core radius as $r_c \sim \xi/\beta_\infty$, which diverges at $\beta_\infty \to 0 $. 

 In the easy-axis phase [$s_0>0$ in Eqs.~ (\ref{Hpol}) and (\ref{HpolA})] there is no magnetic vortices with circular spin currents at large distances. However, skyrmion with the charge $Q= 1$ ($\beta_\infty =\pi$) is still possible and  shown in Fig.~\ref{Fig3}. Without anisotropy [$\xi \to \infty $ in \eq{beta}] spatial rescaling does not change the energy of the $Q= 1$ skyrmion and it can have any size.  At fixed scale-invariant distribution of $\beta$ the easy-axis anisotropy energy is smaller at smaller skyrmion size, and the skyrmion is expected to collapse to very small size. But anisotropy modifies this distribution, and this could stabilize the skyrmion at the scale $\xi$.  Numerical calculation \cite{Sp1} showed that the vortex with the  $Q= 1$ skyrmion core is unstable in  ferromagnetic insulators 
 discussed in the present section, but is stable in the ferromagnetic spin-1 BEC (see below). In ferromagnetic insulators the $Q= 1$ skyrmion  can be stabilized by other interactions, e.g., by gradient terms of higher order \cite{Ivan,Aban}, or by the Dzyaloshinskii--Moriya interaction \cite{Nagaos}. 

We consider the quasimomentum balance in the coordinate frame moving with constant vortex velocity $\bm v_L$, calculating the total quasimomentum flux through cylindrical surfaces restricting the area around the vortex. All time derivatives are $\partial /\partial t=-(\bm v_L\cdot \bm \nabla)$, and the quasimomentum flux tensor in the moving frame is   
\be
\Pi_{ij}={\hbar ^2 \rho_2 \over 2 m^2}\left(\nabla_i \beta \nabla_j \beta+\sin^2 \beta  \nabla_i \varphi \nabla_j \varphi\right)-g_iv_{Lj}+P\delta_{ij},
    \ee{}
where the quasimomentum density is
\be
\bm g = - {\partial {\cal L} \over \partial \dot\varphi} \bm \nabla\varphi =-  {\hbar \over m} \rho_1 \cos \beta  \bm \nabla\varphi.
   \ee{QuM}
The quasipressure $P$ is equal to the Lagrangian (\ref{LagMl}).  We  expand the expressions for $P$ and $\Pi_{ij}$ in small deviations  $\beta'$  and $\bm \nabla \varphi'$ from values of  the polar angle $\beta$  and the gradient 
 $\bm \nabla \varphi$ in  the stationary  vortex. Only terms proportional to $\bm v_L$ and to the constant phase gradient deviation $\bm \nabla \varphi'$ at large distances from the vortex axis, which is connected with the spin current past the vortex, 
 \be 
\bm j_z= -{\hbar^2 \rho_2\over 2m^2}\sin^2\beta_\infty  \bm \nabla\varphi',
    \ee{}
 are important  for the quasimomentum balance. Correction to
the quasimomentum flux is
\be
\Pi'_{ij}= {\hbar ^2 \rho_2 \over 2 m^2}\sin^2 \beta  \nabla_i \varphi \nabla_j \varphi'-v_{Lj}g_i+P'\delta_{ij} 
= {\hbar ^2 \rho_2 \over 2 m^2}\sin^2 \beta  \nabla_i \varphi \nabla_j \varphi'+ {\hbar  \rho_1 \over  m}\cos\beta \nabla_i\varphi v_{Lj}+P'\delta_{ij},
    \ee{}
where 
the quazipressure perturbation is 
\be
P'= -   {\hbar\rho_1 \over  m}\cos \beta(\bm v_L \cdot \bm \nabla)  \varphi - {\hbar ^2 \rho_2 \over  2m^2}\sin^2\beta (\bm \nabla \varphi \cdot \bm \nabla \varphi').
    \ee{}
In the LLG theory the quasimomentum density $\bm g$ given by \eq{QuM} diverges at the axis of the vortex. Because of it the total variation of the quasimomentum in the area around the vortex is determined by the quasimomentum flux   not only through  the surface  at large distances $r$ but also through  the surface  at very small distances $r$ from the vortex axis \cite{NikSon}:
\be
\int\limits_{r\to \infty } \Pi_{ij} dS_j- \int\limits_{r\to 0 } \Pi_{ij} dS_j =-{2\pi\hbar    \over  m}\left\{\rho_1 (1-\cos\beta_\infty)[\hat z \times \bm v_L]_j+{\rho_2\sin^2 \beta_\infty\over 2}[\hat z \times \bm\nabla  \phi']_j\right\}=0.
     \ee{balQu}
The  quasimomentum balance equation is at the same time the force balance equation. The term proportional to the vortex velocity $\bm v_L$  is a gyroscopic force similar to the Magnus force on the hydrodynamic vortex. The second term  in the right-hand side is a force produced by a spin supercurrent past the vortex (analog of the Lorentz force on the vortex in superconductors and superfluids). The Lorentz force  is a gradient of the energy of interaction between the vortex and the spin current past the vortex. This energy is determined by cross terms containing the phase gradient  (\ref{grVor}) induced by the vortex and the phase gradient $\bm\nabla \varphi'$ produced by the spin current. We assumed that the spin current is constant far from the vortex line. But in general it can vary at scales essentially exceeding the vortex core radius, taking into account phase variation induced by other distant vortices. In this case \eq{balQu} contains $\bm\nabla \varphi'$ at distances much larger than the vortex core radius, but much smaller than the distance from other vortices.

\Eq{balQu} yields the relation between the vortex velocity $\bm v_L$ and the spin current $\bm j_z$. Up to now we considered the vortex with positive polarization and one $2\pi$-rotation of the spin $\bm s$ around the vortex axis. Generalizing 
for arbitrary polarization $\nu=\pm 1$ and arbitrary integer number $n$ of rotation of $\bm s$:
\be
\bm v_L=-{n(\nu+\cos\beta_\infty)\over 2}{\rho_2\over \rho_1} \bm\nabla  \phi'={nm^2\over \rho_1 \hbar^2(\nu-\cos\beta_\infty)}\bm j_z.
  \ee{}

\section{Dynamics of vortices in the ferromagnetic spin-1 BEC} \label{vdb}

In an axisymmetric vortex with a single quantum of circulation of the spin phase $\varphi$  the    azimuthal velocity around the vortex axis in general is
\be
v_s(r)=  {\hbar[N-\cos\beta(r)] \over mr}.
    \ee{vgam} 
Here the integer $N$ points out the number of full $2\pi$ rotations of the Euler angle $\alpha$ around the vortex axis [see \eq{vEu}].  This velocity satisfies the Mermin--Ho theorem connecting the velocity with variation  of $\bm s$ [the term $\propto \cos\beta(r)$].

Both contributions are singular at $r\to 0$. However, we look only  for nonsingular vortices, with the energy smaller than singular ones. Two singular contributions to the velocity cancel one another if $N=1$ for the vortex with positive polarization and    $N=-1$
 for the vortex with negative polarization. 
 
 Taking into account \eq{vgam}  with $N=1$ (positive polarization)  the Hamiltonian (\ref{Hpol}) for axisymmetric vortex becomes
\bem
 H=  \rho  \left\{{\hbar ^2 \over  4m^2}\left[{(2-\cos\beta)^2-1 \over r^2} +\left(d \beta\over dr\right)^2\right]
+{G(\cos\beta -s_0)^2\over 2}\right\}.
   \eem{HpolH}
The Euler--Lagrange equation for this Hamiltonian is
 \be
{d ^2 \beta \over dr^2}+{1\over r}{d  \beta \over dr}-\sin\beta\left({2- \cos\beta\over r^2} 
-{\cos\beta -s_0 \over \xi^2}\right)=0,
         \ee{betaH}
where $s_0=\cos\beta_\infty<1$ in the easy-plane phase.  In the theory  of the A phase of superfluid $^3$He the vortex at $\beta_\infty =\pi/2$ (meron) was known as Mermin--Ho vortex, while the vortex at $\beta_\infty =\pi$ was called the Anderson--Toulouse  vortex \cite{Sal85}. This vortex has circulation of $\bm v_s$ but  no circulating spin current far from the vortex. All other vortices at $\beta_\infty <\pi$ have both circulations.  In contrast to ferromagnetic insulators with the Euler--Lagrange equation (\ref{beta}), according to numerical solution of \eq{betaH}  \cite{Sp1}, anisotropy is able to stabilize the skyrmion with the charge 1 ($\beta_\infty =\pi$) [see discussion below \eq{beta1}].

The ferromagnetic spin-1 BEC is Galilean invariant, and the quasimomentum does not differ from true momentum. As in the previous section, we consider  the momentum  balance  in the coordinate frame moving with the velocity $\bm v_L$ and expand the momentum flux tensor $\Pi_{ij}$  in small perturbations produced by vortex motion and currents past the vortex:
\bem
\Pi'_{ij}=\rho v_{si}(v_j-v_{Lj})+  {\hbar ^2 \rho \over 2 m^2}\sin^2 \beta  \nabla_i \varphi \nabla_j \varphi'+P'\delta_{ij}, 
    \eem{}
where the pressure perturbation is determined from the Bernoulli law:
\be
P'={\hbar\rho \over  m}[(\bm v_L-\bm v_s) \cdot \bm \nabla]  \theta- {\hbar ^2 \rho \over  2m^2}\sin^2\beta \nabla \varphi \nabla \varphi'  .
  \ee{}
 In contrast to the LLG theory for localized spins, there is no terms in the momentum flux tensor divergent at the vortex axis. Therefore the variation of the total momentum around the vortex is determined only by the momentum flux through the surface far away from the vortex axis:
 \be
  \int\limits_{r\to \infty } \Pi_{ij} dS_j ={2\pi\hbar  \rho  \over  m}\left\{ (1-\cos\beta_\infty)[\hat z \times (\bm v_s- \bm v_L)]_j-{\sin^2 \beta_\infty\over 2}[\hat z \times \bm\nabla  \phi']_j\right\}=0.
     \ee{balMo}
Generalizing this equation on the vortex  with polarization $\nu=\pm 1$ and integer number $n$ of $2\pi$-rotations of $\bm s$ around the $z$ axis the vortex velocity  is
\be
\bm v_L =\bm v_s-{n(\nu+\cos\beta_\infty)\over 2}\bm\nabla  \phi' = {\bm j\over \rho}+ {nm\over \hbar \rho(\nu-\cos\beta_\infty) }\bm j_z.
    \ee{}
This is a generalization of Helmholtz's theorem, which tells that in a scalar superfluid (or an ideal fluid in classical hydrodynamics) the vortex moves with the fluid velocity $\bm v_s$. In the ferromagnetic spin-1 BEC not only the mass current $\bm j$ but also spin current $\bm j_z$ produces the Lorentz force driving the vortex. The transverse gyroscopic force (Magnus force in hydrodynamics) is 
\be 
\bm F_G=-{2\pi n\hbar  \over  m} (\nu-\cos\beta_\infty) \rho [\hat z \times  \bm v_L].
   \ee{}
The force is proportional to the circulation of the superfluid velocity $\oint \bm v_s\cdot d\bm l =(nh/m)(\nu-\cos\beta_\infty)$. It is interesting that the gyrotropic force  in ferromagnetic insulator [see \eq{balQu}] is given by a similar expression (apart from the difference between two densities $\rho$ and $\rho_1$), although there is no superfluid velocity in the theory.   On the other hand, the superfluid circulation is proportional to the skyrmion charge in the vortex core, which is present in the both theories. Thus, a more careful statement is that the gyroscopic force is proportional to the core skyrmion charge.

\section{Discussion and conclusions}  \label{dis}

We analyzed dynamics of nonsingular vortices  in a ferromagnetic spin-1 BEC, where both mass and spin superfluidity are possible in the presence of uniaxial anisotropy. Vortices are nonsingular only if there is circulation of the wave function phase and the spin phase (the angle of spin rotation around a chosen axis). Their cores have structure of skyrmions with charges tuned by uniaxial anisotropy.

The equation of vortex motion is derived from the quasimomentum conservation law following from Noether's theorem for translationally invariant  media. The ferromagnetic spin-1 BEC is Galilean invariant, and the quasimomentum does not differ from the true momentum.  Vortex dynamics   in a ferromagnetic spin-1 BEC is compared with dynamics of magnetic vortices following from the  LLG theory for ferromagnetic insulators. In the latter case the vortex is driven by the spin current past the vortex, while in the former one both the mass and the spin currents make the vortex to move. In the both cases the driving force (Lorentz force in superfluid hydrodynamics) is balanced by the transverse gyrotropic force proportional to the vortex velocity $\bm v_L$ (analog of the Magnus force in scalar superfluids and classical ideal fluids). The gyrotropic force is proportional to a charge of a skyrmion emerging in a vortex core. 

 In a ferromagnetic spin-1 BEC the core skyrmion charge determines circulation of the superfluid velocity. On the other hand, the frequencies of vortex precession  in a potential trap or  of Kelvin waves along vortex lines are proportional to the superfluid circulation \cite{EBS}. This can be used for experimental check of the results of the present analysis. 



%

\end{document}